\documentclass[prd,superscriptaddress,onecolumn,showkeys]{revtex4}
\usepackage{textcomp}
\usepackage{eurosym}
\usepackage{amsfonts}
\usepackage{array}
\usepackage{amsthm}
\usepackage{bm}
\usepackage{palatino}
\usepackage[colorinlistoftodos]{todonotes}
\usepackage{mathpazo}
\usepackage{supertabular}
\usepackage{subfig}
\usepackage{amssymb}
\usepackage{eurosym}
\usepackage{amsmath}
\usepackage{epsfig}
\usepackage{graphics}
\usepackage{changes}
\usepackage{color}
\usepackage{graphicx}
\usepackage[colorlinks=true,
            linkcolor=blue,
           urlcolor=black,
           citecolor=black]{hyperref}

\def\be{\begin{equation}}
\def\ee{\end{equation}}
\def\beq{\begin{eqnarray}}
\def\eeq{\end{eqnarray}}

\def\bes{\begin{eqnarray}}
\def\ees{\end{eqnarray}}

\begin{document}
\title{Gravitational Analysis of Rotating Charged Black Hole-Like Solution in Einstein-Gauss-Bonnet Gravity}

\author{Riasat Ali}
\email{riasatyasin@gmail.com}
\affiliation{Department of Mathematics, GC,
University Faisalabad Layyah Campus, Layyah-31200, Pakistan}

\author{Rimsha Babar}
\email{rimsha.babar10@gmail.com}
\affiliation{Division of Science and Technology, University of Education, Township, Lahore-54590, Pakistan}

\author{Muhammad Asgher}
\email{m.asgher145@gmail.com}
\affiliation{Department of Mathematics, The Islamia
University of Bahawalpur, Bahawalpur-63100, Pakistan}
\begin{abstract}
This work analyzes the Einstein-Gauss-Bonnet gravity of charged black hole 
solutions through Newman-Janis approach. The Hawking temperature for
corresponding  black hole is also computed. The solution depends upon 
rotation parameter $a$, black hole mass, charge and horizon. Moreover, the 
graphical behavior of temperature via event horizon to analyze the stability of
black hole under the effects of rotation parameter is discussed.  The graphs
are plotted in the presence/absence of rotation parameter and charge.
Furthermore, the Hawking temperature under gravity effects is studied by
using the semi-classical method. It is also observed that the maximum
temperature at non-zero horizon depicts the BH remnant. Finally, the
logarithmic corrected entropy for given black hole is computed and the
logarithmic corrected entropy under effects of rotation and correction
parameter are studied.
\end{abstract}


\date{\today}

\maketitle

\section{Introduction}

In order to study the gravity theories of higher-order curvature \cite{1, 2},
a few modifications have been introduced in general relativity (GR) theory,
a well known theory is called Einstein Gauss–Bonnet (EGB) gravity theory.
This theory is actually a rare case of Lovelock gravity and also a modification of GR.
The EBG gravity Lagrangian is a special type of the Lovelock gravity Lagrangian \cite{3}.
One interesting feature of EGB is that it maintains the second-order equation of motion in
self-assertive number of dimensions $D\geq5$. This significant feature gives the premise
to research on gravity in the other structure. Although, because of the metric definition
in dimensions $D>4$, the extent of exploration is restricted to space-time with higher
dimensions. By re-scaling the coupling constant $\alpha$, Glavan and Lin constructed
$4$-dimensions of EGB gravity  \cite{4}. They showed if we take the product of GB action
with a factor $\frac{\alpha}{D-4}$ rather than $\alpha$, then we will consider an unchanged
contribution to Einstein’s equations in dimension $D = 4$. After this presented method,
several investigations about the EGB gravity and its thermodynamics have been done.
Many investigations about the GB gravity as well as its properties have been made
after presented this valuable method \cite{4a,4ab}.
The topological BH solutions of GB gravity have been derived using two kinds of nonlinear electrodynamics \cite{4b} and also discussed that the stability of heat capacity.
Clifton and his colleagues \cite{5} searched for the
observational constraints on regular EGB gravity in $4$ dimensions and find out the range
of coupling constant. The following features about black holes (BHs) i.e., radiation,
instability, shadows, quasinormal modes and grey-body factors in $4$ dimensions EGB gravity
have been studied in \cite{6}-\cite{10}.
The most recent twenty years had a great deal of treat for theoretical cosmologists, coming from both cosmological scale information and furthermore from astrophysical scales occasions. Especially, the perception of the right now speeding up Universe
coming from the standard candles SNe Ia \cite{1x}, has completely changed our discernment of how the Universe evolves. The gravitational wave recognition coming from the neutron star consolidating GW170817 event \cite{2x}, has totally influenced modified speculations of gravity, barring a few of these from being reasonable depictions of our Universe at astrophysical scales. Especially, the GW170817 occasion showed that the proliferating rate of the gravitational waves $c_T$ is equivalent
to unity to that of light, specifically $c^2_T\simeq1$, in standard units. This factor as we specified, avoided quickly numerous elective speculations of gravity. Although, there exist many altered gravity speculations that actually stay strong against the GW170817 event results.
Oikonomou and his colleagues \cite{3x}-\cite{7x} have explored how the EBG theories can be rendered reasonable and viable with the GW170817 event. The gravitational wave speed for an EBG theory is equal to,
$c^2_T = 1-\frac{Q_f}{2Q_t}$
 with $Q_f=8(\ddot{\xi}- H\dot{\xi})c_1$, while $c_1$  represents the dimensionless constant multiplication factor that work in the Lagrangian with GB term, and the function $Q_t$ is 
a function of $H$ where $H$ stands for Hubble rate. So as indicated by our supposition, if $Q_f=0$ the gravitational wave speed would be equivalent to one. This infers that the GB coupling must holds the differential condition $\ddot{\xi}- H\dot{\xi}=0$, which suggests that $\dot{\xi} = e^N$ , here $N$ denotes the e-foldings number.
 
This technique empowered us to communicate all the slow-roll as well as the observational indices as elements of the e-foldings number, and in the long run concentrate on the phenomenology of the demonstrate.

Semiclassical techniques based on Hawking radiation as a tunneling impact were introduced
in the course of the last two decade and have gained a great deal of interest. This approach
has been presented by Hawking at the point when he proposed his interesting theoretical disclosure
named "Hawking radiation" \cite{11} and after this, it has been clarified by Parikh and Wilczek
\cite{12, 13} that how Hawking radiation occurs. The physical significance of this emission action shows that vacuum fluctuations accelerate particle (positive mass) anti-particle (negative mass) pairs towards the BH horizon. Hawking studied that positive mass particles have enough energy to radiate from the BH, while negative mass particles have no capability to emit from the BH.
Parikh
and Wilczek provided a complete path to this theoretical discovery in the form of WKB approximation.
This method utilize geometric optic estimation which is another perspective of eikonal estimation in
wave explanation  \cite{14}. Particles are set in front of a boundary, when particles tunnel from this
boundary the BH mass reduces in the form of particles energy. In spite of the fact that with the
Parikh-Wilczek approach, a more precise strategy of the tunneling was given, but there were as yet
unsolved issues like information loss, temperature divergence and unitary. Numerous  investigations
have been done on the radiation and tunneling approach from the different BH horizons; Some of these
most significant investigations can be seen in \cite{15}-\cite{45}. The tunneling radiation for diferrent BHs have been studied and also examined the  tunneling radiation with the effects of the BH geometry and various parameter's. It is possible to discuss quantum-corrected thermodynamical features of BH by integrating generalised uncertainty principle (GUP) effects \cite{45a}. The GUP gives high-energy solution to BH thermodynamics, allowing quantum gravity theory to have a minimal length. By taking into account the GUP effects, it is very feasible to study the quantum modified thermodynamics of BHs.
The GUP relation satisfies the following expression \cite{40, 45b}
\begin{equation*}
\Delta x\Delta p \geq  \frac{1}{2}\Big(1+\alpha(\Delta p)^2\Big).
\end{equation*}
here $\alpha=\frac{\alpha_{\circ}}{M^{2}_{p}}$, $M^{2}_{p}$ shows the Plank's mass
and $\alpha_o<10^{34}$ represents the dimensionless parameter.

The main purpose of this paper is to investigate the charged BH in EGB gravity for rotating case
and then to discuss its thermodynamics i.e., Hawking temperature. Moreover, to check
the stability condition of charged rotating BH in EGB gravity by graphical interpretation of Hawking temperature with horizon.
In order to meet our aim, we attempt to construct the rotating charged BH solutions of EBG
gravity with the help of Newman-Janis algorithm and then
investigate their Hawking temperature and study the effects of rotation parameter with the help
of plots. Furthermore, by using the quantum tunneling approach, we also study their quantum
corrected Hawking temperature by using the semi-classical method with the help of WKB approximation.
At last, we analyze the gravity effects by plotting the graphs of corrected temperature with horizon,
which depicts the stability of BH in rotating case.
This article is arranged as follows, Sec. \textbf{II} contains the spaec-time information of $D$-dimensional BHs. Subsection \textbf{II. A}, investigate the solution
of rotating charged BH in EBG gravity by using the Newman-Janis method and study the temperature
for the associated BH. Sec. \textbf{III}, gives the graphical temperature behaviour w.r.t event horizon
and analyze the effects of rotation parameter. Sec. \textbf{IV}, check the quantum corrected temperature
for EGB gravity of charged rotating BH by using quantum tunneling method for boson particles in $4$
dimension spaces.
In Sec. \textbf{V}, we study the quantum gravity graphical effects on
corrected Hawking temperature for boson particles. Section \textbf{VI} describes the logarithmic corrected entropy for EGB gravity of rotating charged BH.
In Sec. \textbf{VII}, we summarize the result of our work.

\section{Einstein Gauss Bonnet Gravity Theory of Black Holes}

In this section, we examine the Einstein-Gauss-Bonnet $D$-dimension charged BHs gravity theory.
To do so, we consider the EGB BHs gravity theory with Lagrangian equation,
\begin{equation}
\L=(R^2+R_{abcd}R^{abcd}\hat{\beta}-4R_{ab}R^{ab})+R.\nonumber
\end{equation}
The solution for
spherically symmetric charged BH with this gravity theory in $D$-dimension space-time can be given as \cite{46}
\begin{equation}
ds^{2}=-G(r)dt^{2}+G^{-1}(r)dr^{2}+r^2d{\Omega}^2_{d-2},\label{aa}
\end{equation}
where
\begin{equation}
G(r)=1+\left[1-\sqrt{1+\frac{16\beta M}{r^{d-1}}-\frac{8\beta q^2}{r^{2d-4}}}\right]\frac{r^2}{4\beta}, \nonumber
\end{equation}
where $M$, $q$ and $\beta$ are BH mass, BH charge and the BH coupling constant $\hat{\beta}$ as $\beta=(d-3)(d-4)\hat{\beta}/2$, respectively.
We study the space-time by $d{\Omega}^2_{d-2}$ of the unit $(d-2)$ area of sphere $A_{d-2}$.
The only nonzero component of the electromagnetic potential has the following form
\begin{equation}
A_{t}(r)=-\frac{q}{r^{d-3}}.
\end{equation}
Here, $M$ and $q$  are associated to the BH mass $\hat{M}$ and  Arnovwitt-Deser-Misner
BH charge$(\hat{q})$ as
\begin{equation}
q^2=\frac{2\hat{q}^2(d-3)}{d-2},~~~~~~~~
M=\frac{8\pi}{A_{d-2}(d-2)}\hat{M}.\nonumber
\end{equation}
The horizon solution at $r=r_h$ can be obtained from the zeros of the function by setting $G(r_h)=0$
\begin{equation}
r_{h}^{2(d-3)}+2\beta r_{h}^{2(d-4)}-2M r_{h}^{d-3}+q^2=0.
\end{equation}
The above equation gives the two real root ($r_h=r_{\pm}$), the positive sign denotes
the largest root as well as negative sign describes the lowest root.
\subsection{Hawking Temperature for Rotating Charged Black Hole}

The spherically symmetric of charged BH solution in $4$-dimensional space-time has the following form
\begin{equation}
ds^{2}=-G(r)dt^{2}+G^{-1}(r)dr^{2}+r^2d\theta^{2}+r^2\sin^2\theta d\psi.\label{ab}
\end{equation}

By the coordinates transformation $(t, r, \theta, \phi)$ to $(u, r, \theta, \phi)$, we have
\begin{eqnarray}
du=dt-\frac{dr}{G(r)}.\label{A}
\end{eqnarray}
After applying the transformation to the metric Eq. (\ref{ab}), we get
\begin{equation}
ds^{2}=-G(r)du^2-2dudr+r^2 d\theta^2+r^2 \sin^2d\phi^2.
\end{equation}
The metric in the null (EF) frame can be expressed as
\begin{eqnarray}
g^{\mu\nu}=-l^\nu n^\mu-l^\mu n^\nu+m^\nu \bar{m}^{\mu}+m^\mu \bar{m}^{\nu}.
\end{eqnarray}
The corresponding components are given as
\begin{eqnarray}
l^{\mu}&=&\delta_{r}^{\mu},~~~n^{\nu}=\delta_{u}^{\mu}-\frac{1}{2} G \delta_{r}^{\mu},\nonumber\\
m^{\mu}&=&\frac{1}{\sqrt{2}r} \delta_{\theta}^{\mu}+\frac{i}{\sqrt{2}r \sin\theta}\delta_{\phi}^{\mu},\nonumber\\
\bar{m}^{\mu}&=&\frac{1}{\sqrt{2}r} \delta_{\theta}^{\mu}-\frac{i}{\sqrt{2}r\sin\theta}\delta_{\phi}^{\mu}.\nonumber
\end{eqnarray}
For all point in the BH metric the null vectors of the null tetrad satisfy the conditions of
$l_{\mu}l^{\mu}=n_{\mu}n^{\mu}=m_{\mu}m^{\mu}=l_{\mu}m^{\mu}=m_{\mu}m^{\mu}=0$
and $l_{\mu}n^{\nu}=-m_{\mu}\bar{m}^{\mu}=1$ in the place of $(u, r)$ coordinate transformation are
$u\rightarrow u-ia \cos\theta$, $r\rightarrow r+ia \cos\theta$, then we perform the transformation
$G(r)\rightarrow \tilde{G}(r, a, \theta)$ and $\Sigma^2=a^2 \cos^2\theta+r^2$.
The null vectors in the $(u, r)$ space can be written as
\begin{eqnarray}
l^{\mu}&=&\delta_{r}^{\mu},~~~n^{\nu}=\delta_{u}^{\mu}-\frac{1}{2}
\tilde{G} \delta_{r}^{\mu},\nonumber\\
m^{\mu}&=&\frac{1}{\sqrt{2}\Sigma}\left(\delta_{\theta}^{\mu}+ia \sin\theta\left(\delta_{u}^{\mu}
-\delta_{r}^{\mu}\right)+\frac{i}{\sin\theta}\delta_{\phi}^{\mu}\right),\nonumber\\
\bar{m}^{\mu}&=&\frac{1}{\sqrt{2}\Sigma}\left(\delta_{\theta}^{\mu}-ia \sin\theta\left(\delta_{u}^{\mu}
-\delta_{r}^{\mu}\right)-\frac{i}{\sin\theta}\delta_{\phi}^{\mu}\right),\nonumber
\end{eqnarray}
From the null tetrad definition of the metric tensor $g^{\mu r}$ are given by
\begin{eqnarray}
g^{uu}&=&\frac{a^2\sin\theta^2}{\sum^2},~~~g^{ur}=g^{ru}=-1-\frac{a^2\sin\theta^2}
{\sum^2},~~~g^{rr}=\tilde{G}+\frac{a^2\sin\theta^2}{\sum^2},~~~
g^{\theta\theta}=\frac{1}{\sum^2},\nonumber\\
g^{\phi\phi}&=&\frac{1}{\sum^2\sin\theta^2},~~g^{u\phi}=g^{\phi u}=\frac{a}{\sum^2}~~~~,
g^{r\phi}=g^{\phi r}=-\frac{a}{\sum^2}.\nonumber
\end{eqnarray}
The new metric according to null tetrad can be obtained as
\begin{equation}
ds^{2}=-\tilde{G}(r)du^2-2dudr-2a \sin\theta^2\Big(1-\tilde{G}\Big)dud\phi+2a
\sin\theta^2drd\phi+\Sigma^2d\theta^2
+\sin\theta^2\left(\Sigma^2-a^2\sin\theta^2\Big(\tilde{G}-2\Big)\right)d\phi^2.
\end{equation}
Now, we consider the coordinate transformation from the Eddington Finkelstein (EF) to Boyer Lindquist (BL) coordinates as
\begin{equation}
du=dt+y(r)dr,~~~d\phi=d\phi+F(r)dr,
\end{equation}
where
\begin{equation}
y(r)=\frac{a^2+r^2}{r^2 \tilde{G} +a^2},~~~~~ F(r)=
-\frac{a}{r^2 \tilde{G} +a^2},~~~~\tilde{G}(r, \theta)=\frac{a^2 \cos^2\theta+r^2 G(r)}{\Sigma^2}.
\end{equation}
Finally, we get the BH spacetime in the form
\begin{eqnarray}
ds^{2}&=&-\left[1+\frac{\frac{r^4}{4\beta}\Big(1-\sqrt{1-\frac{8\beta q^2}{r^{4}}+\frac{16\beta M}{r^{3}}}\Big)}{\Sigma^2}\right]dt^2+\frac{\Sigma^2}{\Delta_r}dr^2-2a \sin^2\theta
\left[\frac{\frac{r^4}{4\beta}\Big(1-\sqrt{1-\frac{8\beta q^2}{r^{4}}+\frac{16\beta M}
{r^{3}}}\Big)}{\Sigma^2}\right]dt d\phi\nonumber\\
&+&\Sigma^2d\theta^2+a^2\sin^2\theta\left[\Sigma^2-a^2\sin^2\theta\frac{\frac{r^4}{4\beta}
\Big(1-\sqrt{1-\frac{8\beta q^2}{r^{4}}+\frac{16\beta M}{r^{3}}}\Big)}{\Sigma^2}\right]d\phi^2,\label{ccc}
\end{eqnarray}
here
\begin{equation}
\Delta_r=r^2+a^2+\frac{r^4}{4\beta}\Big(1-\sqrt{1-\frac{8\beta q^2}{r^{4}}+\frac{16\beta M}{r^{3}}}\Big).\nonumber
\end{equation}
Now, we discuss the thermodynamical property (i.e., Hawking temperature) for corresponding BH.
In order to derive the Hawking temperature, we use the following formula for spherically symmetric BHs
\begin{equation}
T_{H} = \frac{\tilde{G}'(r_+)}{4\pi}.
\end{equation}
The Hawking temperature for
\begin{eqnarray}
T_{H}&=&\frac{r_{+}^5\Big(\Upsilon-r_{+}^2\Big)-4Mr_{+}^4\beta+2a^2r_{+}\left(4q^2
\beta+r_{+}^4\Big(\Upsilon-r_{+}^2\Big)-10Mr_{+}\beta\right)}{8\pi \Upsilon\beta\Big(r_{+}^2+a^2\Big)},
\end{eqnarray}
where
\begin{equation*}
\Upsilon=\sqrt{r_{+}^2-8q^2\beta+16Mr_{+}\beta}.
\end{equation*}
The Hawking temperature depends upon the BH mass $M$, arbitrary constant $\beta$,
BH charge $q$, spin parameter $a$ and BH horizon radius $r_{+}$.

\section{Stability Analysis Via Graphical Interpretation}

This section investigate the stability of charged rotating BH with EBG gravity in the
presence and absence of spin parameter $a$. We observe the horizon structure of BH by
plotting the graphs of $\Delta_r$ with respect to horizon $r_+$.
as well as discuss the stability condition for BH under the effects of different parameters i.e.,
BH charge $q$ and rotation parameter $a$ in the range $0\leq r_+\leq15$.

Fig. \textbf{1}. shows the behavior of $\Delta_r$ with $r_+$ for fixed
value of $M=1$ (BH mass) and $\beta=0.01$ (arbitrary parameter) in the presence and absence of spin parameter.
In (i) the different values of spin parameter $a$ for fixed values of BH charge $q=1$
satisfy the relation of $\Delta_r$ with horizon equation by showing the positive behavior.
In (ii) one can observe that in the absence of rotation parameter $a=0$ and for different values
of charge $q$, the behavior also satisfies the horizon relation.
\begin{center}
\includegraphics[width=6.5cm]{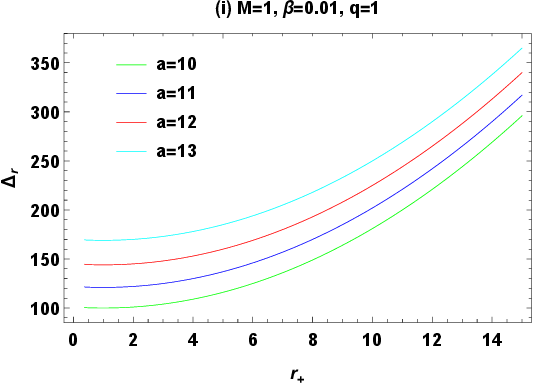}\includegraphics[width=6.5cm]{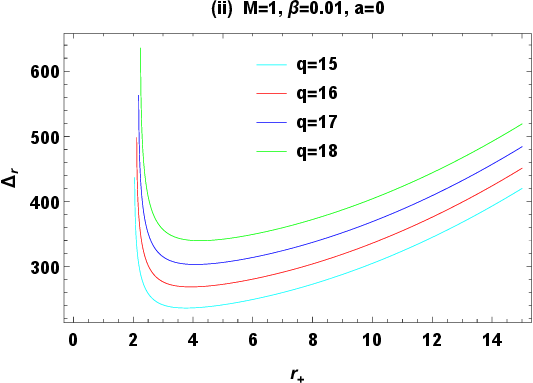}\\
{Figure 1: $\Delta_{r}$ via horizon $r_{+}$ for $M=1$, $\beta=0.01$. Left $q=1$, $a=10$ (solid green), $a=11$ (solid blue), $a=12$ (solid red), $a=13$ (solid cyan). Right $a=0$, $q=15$ (solid cyan), $q=16$ (solid red), $q=17$ (solid blue), $q=18$ (solid green).}
\end{center}

\begin{center}
\includegraphics[width=6.5cm]{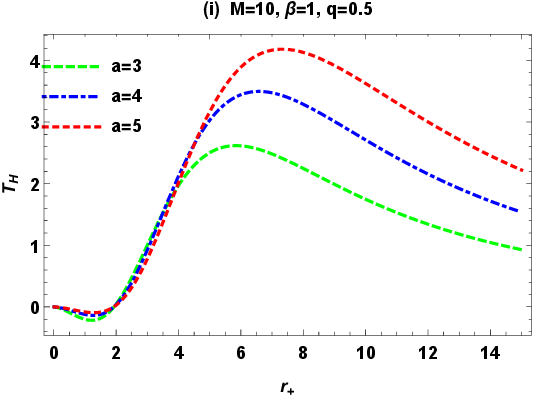}\includegraphics[width=6.7cm]{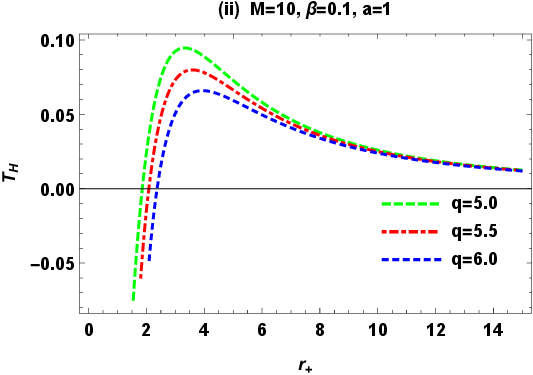}\\
{Figure 2: $T_{H}$ via $r_{+}$ for $M=10$. Left $\beta=1$, $q=0.5$, $a=3$ (dashed), $a=4$ (dot-dashed), $a=5$ (small-dashed). Right $\beta=0.1$, $a=1$, $q=5$ (dashed), $q=5.5$ (dot-dashed), $q=6$ (small-dashed).}
\end{center}
Fig. \textbf{2}. represents the temperature $T_H$ behavior w.r.t
horizon $r_+$ for fixed value of BH mass $M=10$.
In (i) the $T_H$ shows the behavior for fixed values of BH charge $q=0.5$,
arbitrary parameter $\beta=1$ and different values of rotation parameter $a$.
One can observe that the temperature slowly goes on increasing with the
increasing values of horizon but after a height it gradually goes on
decreasing with the increasing horizon which satisfies the Hawking's
phenomenon, so grantee's the physical and stable form of BH. One can also observe that
with the increasing value of spin parameter the values of temperature increases.

The plot (ii) depicts the behavior of $T_H$ for constants values of spin parameter $a=1$, arbitrary parameter $\beta=0.1$ and different values of charge $q$.
We can observe at initial the BH is not stable but as time goes on BH attains its stable form,
when it eventually drops down from a height and obtains an asymptotically flat state
till $r_+\rightarrow\infty$. The decreasing temperature with increasing horizon also shows the stable condition of BH.
One can observe with the increasing values of charge the values of temperature decreases.

Fig. \textbf{3}. actually represents the behavior of temperature with event horizon
for special case in the absence of spin parameter $a$ or charge $q$.
(i) shows the behavior of $T_H$ for constant values of mass $M=15$, arbitrary parameter
$\beta=0.01$ and different values of $q$ (charge) in the absence of $a$ (rotation parameter).
One can analyze in the absence of rotation BH parameter also shows its stable form after passing some time.
From Fig. \textbf{2} and \textbf{3}, we can observe in the presence of rotation
BH parameter the temperature increases as compare in the absence of $a$.

(ii) depicts the $T_H$ behavior for changing values of rotation BH parameter $a$ and fixed the parameter
 of $\beta=1,~M=25$ in the absence of charge $q$. One can observe that in the absence of
charge the BH is also in its stable form. We can also see that as compare to Fig. \textbf{2} (ii) the
temperature is higher in Fig. \textbf{3} (ii) in the absence of charge.
\begin{center}
\includegraphics[width=7.5cm]{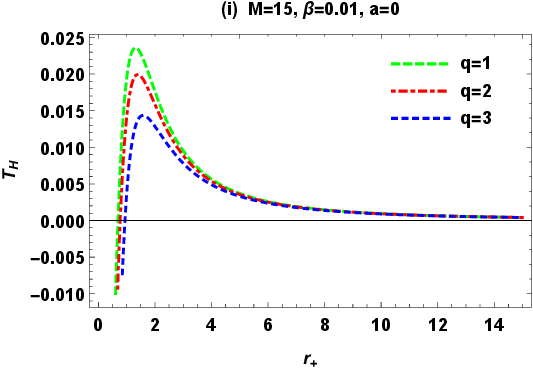}\includegraphics[width=7.1cm]{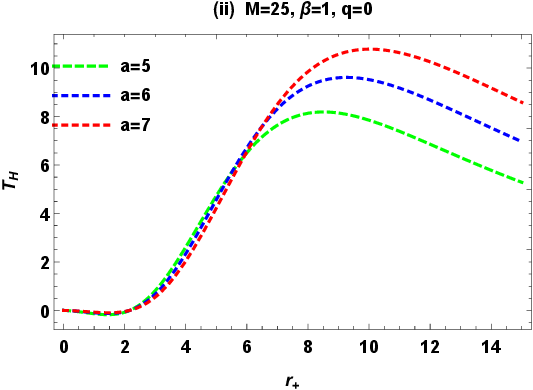}\\
{Figure 3: $T_{H}$ via $r_{+}$. Left $M=15$, $\beta=0.01$, $a=0$, $q=1$ (dashed), $q=2$ (dot-dashed), $q=3$ (small-dashed). Right $M=25$,$\beta=1$, $q=0$, $a=5$ (dashed), $a=6$ (dot-dashed), $a=7$ (small-dashed). }
\end{center}

\section{Corrected tunneling}

The metric Eq. (\ref{ccc}) can be re-expressed as

\begin{eqnarray}
ds^{2}&=&-\tilde{G}dt^{2}+\tilde{H}dr^{2}+\tilde{X}d\theta^{2}
+\tilde{Y} d\phi^{2}+2\tilde{Z}dtd\phi,\label{aa1}
\end{eqnarray}
where
\begin{eqnarray}
\tilde{G}&=&\left[1+\frac{\frac{r^4}{4\beta}\Big(1-\sqrt{1-\frac{8\beta q^2}{r^{4}}
+\frac{16\beta M}{r^{3}}}\Big)}{\Sigma^2}\right],~~\tilde{Y}=
a^2\sin^2\theta\left[\Sigma^2-a^2\sin^2\theta\frac{\frac{r^4}{4\beta}
\Big(1-\sqrt{1-\frac{8\beta q^2}{r^{4}}+\frac{16\beta M}{r^{3}}}\Big)}{\Sigma^2}\right],\nonumber\\
\tilde{Z}&=&a \sin^2\theta
\left[\frac{\frac{r^4}{4\beta}\Big(\sqrt{1-\frac{8\beta q^2}{r^{4}}+\frac{16\beta M}
{r^{3}}}\Big)-1}{\Sigma^2}\right],~~~\tilde{H}=\frac{\Sigma^2}{\Delta_r},~~~
\tilde{X}=\Sigma^2.\nonumber
\end{eqnarray}
The modified filed equation is expressed by \cite{40, 44}
\begin{equation}
\partial_{\mu}\Big(\sqrt{-g}\phi^{\nu\mu}\Big)+\sqrt{-g}\frac{m^2}{\hbar^2}\phi^{\nu}+\sqrt{-g}\frac{i}{\hbar}A_{\mu}\phi^{\nu\mu}
+\sqrt{-g}\frac{i}{\hbar}eF^{\nu\mu}\phi_{\mu}+\hbar^{2}\alpha\partial_{0}\partial_{0}\partial_{0}\Big(\sqrt{-g}g^{00}\phi^{0\nu}\Big)
-\alpha \hbar^{2}\partial_{i}\partial_{i}\partial_{i}\Big(\sqrt{-g}g^{ii}\phi^{i\nu}\Big)=0,\label{bbb}
\end{equation}
here $\phi^{\nu\mu}$, $m$ and $g$ represent the anti-symmetric tensor, mass of particle and coefficient matrix of its determinant, since
\begin{equation}
\phi_{\nu\mu}=\Big(1-\alpha{\hbar^2\partial_{\nu}^2}\Big)\partial_{\nu}\phi_{\mu}-
\Big(1-\alpha{\hbar^2\partial_{\mu}^2}\Big)\partial_{\mu}\phi_{\nu}
+\Big(1-\alpha{\hbar^2\partial_{\nu}^2}\Big)\frac{i}{\hbar}eA_{\nu}\phi_{\mu}
-\Big(1-\alpha{\hbar^2}\partial_{\nu}^2\Big)\frac{i}{\hbar}eA_{\mu}\phi_{\nu},~~\textmd{and}~~
F_{\nu\mu}=\nabla_{\nu} A_{\mu}-\nabla_{\mu} A_{\nu},\nonumber
\end{equation}
where $\alpha,~e~,~A_{\mu}$ and $\nabla_{\mu}$ are the correction parameter, particle charge
BH vector potential and covariant derivative, respectively.
The anti-symmetric tensor with its non-zero components can be derived in the form
\begin{eqnarray}
\phi^{0}&=&\frac{-\tilde{Y}\phi_{0}+\tilde{Z}\phi_{3}}{\tilde{G}\tilde{Y}+\tilde{Z}^2},~~~\phi^{1}=\tilde{G}\phi_{1},
~~~\phi^{2}=\frac{1}{\tilde{X}}\phi_{2},~~~
\phi^{3}=\frac{\tilde{Z}\phi_{0}+\tilde{G}\phi_{3}}{\tilde{G}\tilde{Y}+\tilde{Z}^2},
~~~\phi^{01}=\frac{\tilde{G}\Big(-\tilde{D}\phi_{01}
+\tilde{Z}\phi_{13}\Big)}{\Big(\tilde{G}\tilde{Y}+\tilde{Z}^2\Big)},\nonumber\\
\phi^{02}&=&\frac{-\tilde{Y}\phi_{02}}{\tilde{X}\Big(\tilde{G}\tilde{Y}+\tilde{Z}^2\Big)},
~~~\phi^{03}=\frac{\Big(-\tilde{G}\tilde{Y}+\tilde{G}^2\Big)\phi_{03}}{\Big(\tilde{G}\tilde{Y}
+\tilde{Z}^2\Big)^2},~~~\phi^{12}=\frac{\tilde{G}}{\tilde{X}}\phi_{12},
~\phi^{13}=\frac{\tilde{G}}{\tilde{G}\tilde{Y}+\tilde{Z}^2}\phi_{13},~~\phi^{23}=\frac{\tilde{Z}\phi_{02}
+\tilde{G}\phi_{23}}{\tilde{X}\Big(\tilde{G}\tilde{Y}+\tilde{Z}^2\Big)}.\nonumber
\end{eqnarray}
The WKB approximation is
\begin{equation}
\phi_{\nu}=k_{\nu}\exp\left[\frac{i}{\hbar}I_{0}(t,r,\phi, \theta)+
\Sigma \hbar^{n}I_{n}(t,r, \phi, \theta)\right].
\end{equation}
By putting all the components in Eq. (\ref{bbb}), we get the following set of field equations in the form
\begin{eqnarray}
&&\frac{\tilde{G}\tilde{Y}}{\Big(\tilde{G}\tilde{Y}+\tilde{Z}^2\Big)}\left[\Big(\partial_{0}I_{0}\Big)\Big(\partial_{1}I_{0}\Big)k_{1}-\Big(\partial_{1}I_{0}\Big)^{2}k_{0}
-\alpha\Big(\partial_{1}I_{0}\Big)^{4}k_{0}+eA_{0}\Big(\partial_{1}I_{0}\Big)k_{1}+\alpha
\Big(\partial_{0}I_{0}\Big)^{3}\Big(\partial_{1}I_{0}\Big)k_{1}
\right]\nonumber\\
&+&\left.\alpha eA_{0}\Big(\partial_{0}I_{0}\Big)^{2}\Big(\partial_{1}I_{0}\Big)k_{1}\right]
-\frac{\tilde{G}\tilde{Z}}{\Big(\tilde{G}\tilde{Y}+\tilde{Z}^2\Big)}\left[\alpha\Big(\partial_{1}I_{0}\Big)^{4}k_{3}+\Big(\partial_{1}I_{0}\Big)^{2}k_{3}
-\Big(\partial_{1}I_{0}\Big)\Big(\partial_{3}I_{0}\Big)k_{1}-\Big(\partial_{1}I_{0}\Big)\Big(\partial_{3}I_{0}\Big)^{2}\alpha k_{1}\right]\nonumber\\
&+&\frac{\tilde{Y}}{\tilde{X}\Big(\tilde{G}\tilde{Y}+\tilde{Z}^2\Big)}\left[\Big(\partial_{0}I_{0}\Big)\Big(\partial_{2}I_{0}\Big)k_{2}
+\Big(\partial_{0}I_{0}\Big)^3\Big(\partial_{2}I_{0}\Big)\alpha k_{2}\right.
-\Big(\partial_{2}I_{0}\Big)^2k_{0}-\Big(\partial_{2}I_{0}\Big)^{4}\alpha k_{0}+eA_{0}\Big(\partial_{2}I_{0}\Big)k_{2}\nonumber\\
&+&\left. eA_{0}\alpha\Big(\partial_{0}I_{0}\Big)^{2}\Big(\partial_{1}I_{0}\Big)k_{2}\right]
+\frac{\tilde{G}\tilde{Y}}{\Big(\tilde{G}\tilde{Y}+\tilde{Z}^2\Big)^2}\left[\Big(\partial_{0}I_{0}\Big)\Big(\partial_{3}I_{0}\Big)k_{3}
+\Big(\partial_{0}I_{0}\Big)^{3}\Big(\partial_{3}I_{0}\Big)\alpha k_{3}-\Big(\partial_{3}I_{0}\Big)^{2}k_{0}\right.
-\Big(\partial_{3}I_{0}\Big)^{4}\alpha k_{0}\nonumber\\
&+&eA_{0}\Big(\partial_{3}I_{0}\Big)k_{3}+eA_{0}\Big(\partial_{0}I_{0}\Big)^{2}
\left.\Big(\partial_{3}I_{0}\Big)k_{3}\right]-m^2\Big(\frac{\tilde{Y} k_{0}-\tilde{Z k_{3}}}{\tilde{G}\tilde{Y}+\tilde{Z}^2}\Big)=0,\label{www}\\
&-&\frac{\tilde{G}\tilde{Y}}{\Big(\tilde{G}\tilde{Y}+\tilde{Z}^2\Big)}
\left[\Big(\partial_{0}I_{0}\Big)^{2}k_{1}+
\Big(\partial_{0}I_{0}\Big)^{4}\alpha k_{1}-\Big(\partial_{0}I_{0}\Big)\Big(\partial_{1}I_{0}\Big)k_{0}
-\Big(\partial_{0}I_{0}\Big)\Big(\partial_{1}I_{0}\Big)^{3}\alpha k_{0}
+eA_{0}\Big(\partial_{0}I_{0}\Big)k_{1}\right.\nonumber\\
&+&\left.eA_{0}\Big(\partial_{0}I_{0}\Big)^{3}\alpha k_{1}\right]+\frac{\tilde{G}\tilde{Z}}{\Big(\tilde{G}\tilde{Y}+\tilde{Z}^2\Big)}
\left[\Big(\partial_{0}I_{0}\Big)\Big(\partial_{1}I_{0}\Big)k_{3}+
\Big(\partial_{0}I_{0}\Big)\Big(\partial_{1}I_{0}\Big)^{3}\alpha k_{3}-\Big(\partial_{0}I_{0}\Big)\Big(\partial_{3}I_{0}\Big)k_{1}\right.\nonumber\\
&-&\left.\Big(\partial_{0}I_{0}\Big)\Big(\partial_{3}I_{0}\Big)^{3}\alpha k_{1}\right]+\frac{\tilde{G}}{\tilde{X}}
\left[\Big(\partial_{1}I_{0}\Big)\Big(\partial_{2}I_{0}\Big)k_{2}+
\Big(\partial_{1}I_{0}\Big)\Big(\partial_{2}I_{0}\Big)^{3}\alpha k_{2}-\Big(\partial_{2}I_{0}\Big)^{2}k_{1}
-\Big(\partial_{2}I_{0}\Big)^{4}\alpha k_{1}\right]\nonumber\\
&+&\frac{\tilde{G}}{\Big(\tilde{G}\tilde{Y}+\tilde{Z}^2\Big)}\left[
\Big(\partial_{1}I_{0}\Big)\Big(\partial_{3}I_{0}\Big)k_{3}
+\Big(\partial_{1}I_{0}\Big)\Big(\partial_{3}I_{0}\Big)^{3}\alpha k_{3}-\Big(\partial_{3}I_{0}\Big)^{2}k_{1}-\Big(\partial_{3}I_{0}\Big)^{4}\alpha k_{1}\right]-\tilde{G}m^2 k_{1}\nonumber\\
&+&\frac{eA_{0}\tilde{G}\tilde{Y}}{\Big(\tilde{G}\tilde{Y}+\tilde{Z}^2\Big)}\left[\Big(\partial_{0}I_{0}\Big)k_{1}+\Big(\partial_{0}I_{0}\Big)^{3}\alpha k_{1}
-\Big(\partial_{1}I_{0}\Big)k_{0}-\Big(\partial_{1}I_{0}\Big)^{3}\alpha k_{0}
+eA_{0}k_{1}+eA_{0}\Big(\partial_{0}I_{0}\Big)^{2}\alpha k_{1}\right]\nonumber\\
&+&\frac{eA_{0}\tilde{G}\tilde{Z}}{\Big(\tilde{G}\tilde{Y}+\tilde{Z}^2\Big)}
\left[\Big(\partial_{1}I_{0}\Big)k_{3}+\Big(\partial_{1}I_{0}\Big)^{3}\alpha k_{3}
-\Big(\partial_{3}I_{0}\Big)k_{1}-\Big(\partial_{1}I_{0}\Big)^{3}\alpha k_{1}\right]=0,\\
&&\frac{\tilde{Y}}{\tilde{X}\Big(\tilde{G}\tilde{Y}+\tilde{Z}^2\Big)}\left[\Big(\partial_{0}I_{0}\Big)^{2}k_{2}+
\Big(\partial_{0}I_{0}\Big)^{4}\alpha k_{2}-\Big(\partial_{0}I_{0}\Big)\Big(\partial_{2}I_{0}\Big)k_{0}
-\Big(\partial_{0}I_{0}\Big)\Big(\partial_{2}I_{0}\Big)^{3}\alpha k_{0}
+eA_{0}\Big(\partial_{0}I_{0}\Big)k_{2}\right.\nonumber\\
&+&\left.eA_{0}\Big(\partial_{0}I_{0}\Big)^{3}\alpha k_{2}\right]
+\tilde{G}\tilde{X}\left[\Big(\partial_{1}I_{0}\Big)^{2}k_{2}+
\Big(\partial_{1}I_{0}\Big)^{4}\alpha k_{2}-\Big(\partial_{1}I_{0}\Big)
\Big(\partial_{2}I_{0}\Big)k_{1}-\Big(\partial_{1}I_{0}\Big)
\Big(\partial_{2}I_{0}\Big)^{3}\alpha k_{1}\right]-\frac{m^2 k_{2}}{\tilde{X}}\nonumber\\
&-&\frac{\tilde{Z}}{\tilde{X}\Big(\tilde{G}\tilde{Y}
+\tilde{Z}^2\Big)}\left[\Big(\partial_{0}I_{0}\Big)\Big(\partial_{3}I_{0}\Big)k_{2}+\Big(\partial_{0}I_{0}\Big)^{3}\Big(\partial_{3}I_{0}\Big)\alpha k_{2}-\Big(\partial_{0}I_{0}\Big)\Big(\partial_{3}I_{0}\Big)k_{0}
-\Big(\partial_{0}I_{0}\Big)^3\Big(\partial_{3}I_{0}\Big)\alpha k_{0}+eA_{0}\Big(\partial_{3}I_{0}\Big)k_{2}\right.\nonumber\\
&+&\left.eA_{0}\Big(\partial_{3}I_{0}\Big)^{3}\alpha k_{2}\right]
+\frac{\tilde{G}}{\tilde{X}\Big(\tilde{G}\tilde{Y}+\tilde{Z}^2\Big)}
\left[(\partial_{2}I_{0})\Big(\partial_{3}I_{0}\Big)k_{3}+
\Big(\partial_{2}I_{0}\Big)^{3}\Big(\partial_{3}I_{0}\Big)\alpha k_{3}-\Big(\partial_{3}I_{0}\Big)^{2}k_{2}
-\Big(\partial_{3}I_{0}\Big)^{4}\alpha k_{2}\right]\nonumber\\
&+&\frac{eA_{0}\tilde{Y}}{\tilde{X}\Big(\tilde{G}\tilde{Y}+\tilde{Z}^2\Big)}
\left[\Big(\partial_{0}I_{0}\Big)k_{2}+\Big(\partial_{0}I_{0}\Big)^{3}\alpha k_{2}-\Big(\partial_{2}I_{0}\Big)k_{0}
-\Big(\partial_{2}I_{0}\Big)^{3}\alpha k_{0}+eA_{0}k_{2}+ eA_{0}\alpha
\Big(\partial_{0}I_{0}\Big)^{2}k_{2}\right]=0,
\end{eqnarray}
\begin{eqnarray}
&&\frac{\tilde{G}\tilde{Y}-\tilde{G}^2}{\Big(\tilde{G}\tilde{Y}+\tilde{Z}^2\Big)^2}
\left[\Big(\partial_{0}I_{0}\Big)^{2}k_{3}+\Big(\partial_{0}I_{0}\Big)^{4}\alpha k_{3}-\Big(\partial_{0}I_{0}\Big)\Big(\partial_{3}I_{0}\Big)k_{0}
-\Big(\partial_{0}I_{0}\Big)\Big(\partial_{3}I_{0}\Big)^{3}\alpha k_{0}
+{eA_{0}}\Big(\partial_{0}I_{0}\Big)k_3\right.\nonumber\\
&+&\left.eA_{0}\Big(\partial_{0}I_{0}\Big)^{3}\alpha k_{3}\right]
-\frac{\tilde{Y}}{\tilde{X}\Big(\tilde{G}\tilde{Y}+\tilde{Z}^2\Big)}
\left[\Big(\partial_{1}I_{0}\Big)^{2}k_{3}+
\Big(\partial_{1}I_{0}\Big)^{4}\alpha k_{3}
-\Big(\partial_{1}I_{0}\Big)\Big(\partial_{3}I_{0}\Big)k_{1}
-\Big(\partial_{1}I_{0}\Big)\Big(\partial_{3}I_{0}\Big)^{3}\alpha k_{1}\right]\nonumber\\
&-&\frac{\tilde{Z}}{\tilde{X}\Big(\tilde{G}\tilde{Y}+\tilde{Z}^2\Big)}
\left[\Big(\partial_{0}I_{0}\Big)\Big(\partial_{2}I_{0}\Big)k_{2}
+\Big(\partial_{0}I_{0}\Big)^{3}\Big(\partial_{2}I_{0}\Big)\alpha k_{2}-\Big(\partial_{2}I_{0}\Big)^{2}k_{0}
+\Big(\partial_{2}I_{0}\Big)^{4}\alpha k_{0}+{eA_{0}}\Big(\partial_{2}I_{0}\Big)k_2
+eA_{0}\alpha \right.\nonumber\\
&&\left.\Big(\partial_{0}I_{0}\Big)^{2}\Big(\partial_{2}I_{0}\Big)k_{2}\right]
-\frac{eA_{0}\tilde{G}}{\tilde{X}\Big(\tilde{G}\tilde{Y}+\tilde{Z}^2\Big)}
\left[\Big(\partial_{2}I_{0}\Big)^{2}k_{3}+\Big(\partial_{2}I_{0}\Big)^{4}\alpha k_{3}
-\Big(\partial_{2}I_{0}\Big)\Big(\partial_{3}I_{0}\Big)k_{2}
-\Big(\partial_{0}I_{0}\Big)\Big(\partial_{3}I_{0}\Big)^{3}\alpha k_{2}\right]\nonumber\\
&+&\frac{eA_{0}\Big(\tilde{G}\tilde{Y}-\tilde{G}^2\Big)}{\Big(\tilde{G}\tilde{Y}
+\tilde{Z}^2\Big)^2}\left[\Big(\partial_{0}I_{0}\Big)k_{3}+
\Big(\partial_{0}I_{0}\Big)^{3}\alpha k_{3}
-\Big(\partial_{3}I_{0}\Big)k_{0}-\Big(\partial_{3}I_{0}\Big)^{3}\alpha k_{0}+eA_{0}k_{3}
+eA_{0}\Big(\partial_{0}I_{0}\Big)^{2}\alpha k_{3}\right]\nonumber\\
&-&\frac{m^2 \Big(\tilde{Z}k_{0}-\tilde{G}k_{3}\Big)}{\Big(\tilde{G}\tilde{Y}+\tilde{Z}^2\Big)}=0.\label{zzz}
\end{eqnarray}
By utilizing the technique of separation of variables, we obtain
\begin{equation}
I_{0}=-\left(E-\sum_{i=1}^{2}j_{i}{\Omega_{i}}\right)t+W(r)+J\phi+\nu(\theta),\label{vvv}
\end{equation}
where $E$ and $J$ represent the particle energy, the particle angular momentum corresponding angle $\phi$. After using the equation (\ref{vvv})
into equations (\ref{www})-(\ref{zzz}), we get a matrix of order $4\times4$ labeled as "$U$"
\begin{equation*}
U(k_{0},k_{1},k_{2},k_{3})^{T}=0.
\end{equation*}
The non-zero elements are expressed as follows
\begin{eqnarray}
U_{00}&=&\frac{-\tilde{G}\tilde{D}}{(\tilde{G}\tilde{Y}+\tilde{Z}^2)}\left[W_{1}^2+\alpha W_{1}^4\right]-\frac{\tilde{Y}}{\tilde{X}(\tilde{G}\tilde{Y}+\tilde{Z}^2)}[J^2+\alpha J^4]
-\frac{\tilde{G}\tilde{Y}}{\Big(\tilde{G}\tilde{Y}+\tilde{Z}^2\Big)^2}\left[\nu_{1}^2+
\alpha \nu_{1}^4\right]-\frac{m^2 \tilde{Y}}{\Big(\tilde{G}\tilde{Y}+\tilde{Z}^2\Big)},\nonumber\\
U_{01}&=&\frac{-\tilde{G}\tilde{D}}{\Big(\tilde{G}\tilde{Y}+\tilde{Z}^2\Big)}\left[\tilde{E}+\alpha \tilde{E}^3+eA_{0}+\alpha eA_{0}\tilde{E}^2\right]W_{1}+\frac{\tilde{G}\tilde{Z}}{(\tilde{G}\tilde{Y}+\tilde{Z}^2)}+[\nu_{1}+
\alpha \nu_{1}^3],\nonumber\\
U_{02}&=&\frac{-\tilde{D}}{\tilde{X}\Big(\tilde{G}\tilde{Y}+\tilde{Z}^2\Big)}\left[\tilde{E}+\alpha \tilde{E}^3-eA_{0}-\alpha eA_{0}\tilde{E}^2\right]J,~~~~~K_{12}=\frac{\tilde{G}}{\tilde{X}}[W_{1}+\alpha W_{1}^3]J,\nonumber\\
H_{03}&=&\frac{-\tilde{G}\tilde{E}}{\Big(\tilde{G}\tilde{Y}+\tilde{Z}^2\Big)}\left[W_{1}^2+\alpha W_{1}^4\right]- \frac{\tilde{G}\tilde{Y}}{\tilde{X}\Big(\tilde{G}\tilde{Y}+\tilde{Z}^2\Big)^2}\left[\tilde{E}+\alpha \tilde{E}^3
-eA_{0}-\alpha eA_{0}\tilde{E}^2\right]\nu_{1}+\frac{m^2\tilde{Z}}{\Big(\tilde{G}\tilde{Y}+\tilde{Z}^2\Big)^2},\nonumber\\
U_{10}&=&\frac{-\tilde{G}\tilde{D}}{\Big(\tilde{G}\tilde{Y}+\tilde{Z}^2\Big)}\left[\tilde{E}W_{1}+\alpha \tilde{E}W_{1}^3\right]
-\tilde{G}m^2-\frac{eA_{0}\tilde{G}\tilde{Y}}{\Big(\tilde{G}\tilde{Y}+\tilde{Z}^2\Big)}\left[W_{1}+\alpha W_{1}^3\right],\nonumber\\
U_{11}&=&\frac{-\tilde{G}\tilde{D}}{\Big(\tilde{G}\tilde{Y}+\tilde{Z}^2\Big)}\left[\tilde{E}^2+\alpha \tilde{E}^4-eA_{0}\tilde{E}-\alpha eA_{0}\tilde{E}W_{1}^2\right]+\frac{\tilde{G}\tilde{Z}}{\Big(\tilde{G}\tilde{Y}+\tilde{Z}^2\Big)}\left[\nu_{1}+
\alpha \nu_{1}^3\right]\tilde{E}-\frac{\tilde{G}}{\tilde{X}}\left[J^2+\alpha J^4\right]\nonumber\\
&-&\frac{\tilde{G}}{\Big(\tilde{G}\tilde{Y}+\tilde{Z}^2\Big)}\left[\nu_{1}
+\alpha \nu_{1}^3\right]-m^2\tilde{G}-\frac{eA_{0}\tilde{G}\tilde{Y}}{\Big(\tilde{G}\tilde{Y}+\tilde{Z}^2\Big)}
\left[\tilde{E}+\alpha \tilde{E}^3-eA_{0}-\alpha eA_{0}\tilde{E}^2\right]
+\frac{eA_{0}\tilde{G}\tilde{Z}}{(\tilde{G}\tilde{Y}+\tilde{Z}^2)}\left[\nu_{1}+
\alpha \nu_{1}^3\right],\nonumber\\
U_{13}&=&\frac{-\tilde{G}\tilde{E}}{\Big(\tilde{G}\tilde{Y}+\tilde{Z}^2\Big)}\left[W_{1}+\alpha W_{1}^3\right]\tilde{E}+ \frac{\tilde{G}}{\Big(\tilde{G}\tilde{Y}+\tilde{Z}^2\Big)^2}\left[W_{1}+\alpha W_{1}^3\right]\nu_{1}
+\frac{\tilde{G}\tilde{Z}eA_{0}}{\Big(\tilde{G}\tilde{Y}+\tilde{Z}^2\Big)}\left[W_{1}+\alpha W_{1}^3\right],\nonumber
\end{eqnarray}
\begin{eqnarray}
U_{20}&=&\frac{\tilde{Y}}{\tilde{X}\Big(\tilde{G}\tilde{Y}+\tilde{Z}^2\Big)}\left[\tilde{E}J+\alpha \tilde{E}J^3\right]+
\frac{\tilde{Z}}{\tilde{X}\Big(\tilde{G}\tilde{Y}+\tilde{Z}^2\Big)}\left[\tilde{E}+\alpha \tilde{E}^3\nu_{1}^2\right]
-\frac{\tilde{Y}eA_{0}}{\tilde{X}\Big(\tilde{G}\tilde{Y}+\tilde{Z}^2\Big)}\left[J+\alpha J^3\right],\nonumber\\
U_{21}&=&\frac{\tilde{G}}{\tilde{X}}\left[J+\alpha J^3\right]W_{1},~~~~~
U_{23}=\frac{\tilde{G}}{\tilde{X}(\tilde{G}\tilde{Y}+\tilde{Z}^2)}[J+\alpha J^3]\nu_{1},\nonumber\\
U_{22}&=&\frac{\tilde{Y}}{\tilde{X}\Big(\tilde{G}\tilde{Y}+\tilde{Z}^2\Big)}\left[\tilde{E}^2
+\alpha \tilde{E}^4-eA_{0}\tilde{E}-\alpha eA_{0}\tilde{E}\right]
-\tilde{G}\tilde{X}+\frac{\tilde{Z}}{\tilde{X}\Big(\tilde{G}\tilde{Y}+\tilde{Z}^2\Big)}\left[\tilde{E}
+\alpha \tilde{E}^3-eA_{0}-\alpha eA_{0}\tilde{E}^2\right]\nu_{1}\nonumber\\
&-&\frac{m^2}{\tilde{X}}-\frac{\tilde{G}}{\tilde{X}\Big(\tilde{G}\tilde{Y}+\tilde{Z}^2\Big)}\left[\nu_{1}^2+
\alpha \nu_{1}^4\right]-\frac{eA_{0}\tilde{Y}}{\tilde{X}\Big(\tilde{G}\tilde{Y}+\tilde{Z}^2\Big)}
\left[\tilde{E}+\alpha \tilde{E}^3-eA_{0}-\alpha eA_{0}\tilde{E}^2\right],\nonumber\\
U_{30}&=&\frac{\Big(\tilde{G}\tilde{Y}-\tilde{A^2}\Big)}{\Big(\tilde{G}\tilde{Y}
+\tilde{Z}^2\Big)^2}\left[\nu_{1}+\alpha \nu_{1}^3\right]\tilde{E}
+\frac{\tilde{Z}}{\tilde{X}\Big(\tilde{G}\tilde{Y}+\tilde{Z}^2\Big)}\left[J^2+\alpha J^4\right]
-\frac{m^2\tilde{Z}}{\Big(\tilde{G}\tilde{Y}+\tilde{Z}^2\Big)}-\frac{eA_{0}
\Big(\tilde{G}\tilde{Y}-\tilde{A^2}\Big)}{\Big(\tilde{G}\tilde{Y}
+\tilde{Z}^2\Big)^2}\left[\nu_{1}+\alpha \nu_{1}^3\right],\nonumber\\
U_{31}&=&\frac{\tilde{G}}{\Big(\tilde{G}\tilde{Y}+\tilde{Z}^2\Big)}\left[\nu_{1}+\alpha \nu_{1}^3\right]W_{1},~~~~~
U_{32}=\frac{\tilde{Z}}{\tilde{X}\Big(\tilde{G}\tilde{Y}+\tilde{Z}^2\Big)}\left[J+\alpha J^3\right]\tilde{E}+
\frac{\tilde{G}}{\tilde{X}\Big(\tilde{G}\tilde{Y}+\tilde{Z}^2\Big)}\left[\nu_{1}+\alpha \nu_{1}^3\right]J,\nonumber\\
U_{33}&=&\frac{\Big(\tilde{G}\tilde{Y}-\tilde{A^2}\Big)}{\Big(\tilde{G}\tilde{Y}+\tilde{Z}^2\Big)}\left[\tilde{E}^2
+\alpha \tilde{E}^4-eA_{0}\tilde{E}-\alpha eA_{0}\tilde{E}^3\right]
-\frac{\tilde{G}}{\Big(\tilde{G}\tilde{Y}+\tilde{Z}^2\Big)}\left[W_{1}^2+\alpha W_{1}^4\right]
-\frac{\tilde{G}}{\tilde{X}\Big(\tilde{G}\tilde{Y}+\tilde{Z}^2\Big)}\left[J^2+\alpha J^4\right]\nonumber\\
&-&\frac{m^2 \tilde{G}}{\Big(\tilde{G}\tilde{Y}+\tilde{Z}^2\Big)}
-\frac{eA_{0}\Big(\tilde{G}\tilde{Y}-\tilde{A^2}\Big)}{\Big(\tilde{G}\tilde{Y}+\tilde{Z}^2\Big)}\left[\tilde{E}
+\alpha \tilde{E}^3-eA_{0}\tilde{E}^2\right],\nonumber
\end{eqnarray}
where $\tilde{E}=E-\sum_{i=1}^{2}j_{i}{\Omega_{i}},~~J=\partial_{\phi}I_{0}$,
$W_{1}=\partial_{r}{I_{0}}$ and $\nu_{1}=\partial_{\theta}{I_{0}}$.
We take the solution for non-trivial $\Big|\textbf{U}\Big|=0$ and get
\begin{eqnarray}\label{a1}
ImW^{\pm}&=&\pm \int\sqrt{\frac{\Big(\tilde{E}-j\Omega-eA_{0}\Big)^{2}
+X_{1}\left[1+\alpha\frac{X_{2}}{X_{1}}\right]}{\frac{\tilde{G}
\Big(\tilde{G}\tilde{Y}+\tilde{Z}^2\Big)}{\tilde{Y}}}}dr,\nonumber\\
&=&\pm i\pi\frac{\Big(\tilde{E}-j\Omega-A_{0}e\Big)+\Big[1+\alpha\Xi\Big]}{2\kappa(r_{+})},
\end{eqnarray}
where
\begin{eqnarray}
X_{1}&=&\frac{\tilde{Z}}{\tilde{G}(\tilde{G}\tilde{Y}+\tilde{Z}^2)}\Big[\tilde{E}
-eA_{0}\Big]\nu_{1}+\frac{1}{\Big(\tilde{G}\tilde{Y}+\tilde{Z}^2\Big)}\Big[\nu_{1}^2\Big]
-\frac{1}{\tilde{G}}m^2,\nonumber\\
X_{2}&=&\frac{\tilde{Y}}{\tilde{G}\Big(\tilde{G}\tilde{Y}+\tilde{Z}^2\Big)}\Big[\tilde{E}^4
-2eA_{0}\tilde{E}^3+(eA_{0})^2\tilde{E}^2\Big]+\frac{\tilde{Z}}{\tilde{G}\tilde{X}
\Big(\tilde{G}\tilde{Y}+\tilde{Z}^2\Big)}\Big[\tilde{E}^3-eA_{0}\tilde{E}^2\Big]\nu_{1}
-\frac{1}{\Big(\tilde{G}\tilde{Y}+\tilde{Z}^2\Big)}\Big[\nu_{1}^4\Big]-W_{1}^4.\nonumber
\end{eqnarray}
We observe the tunneling radiation without back-reaction and self-gravity interaction and also explore the corrected tunneling probability rate. The tunneling phenomenon is computed under the consideration of charge-energy conservation and quantum gravity effects. The modified tunneling probability depends on the geometry of BHs as well as quantum gravity.
The modified tunneling probability for charged rotating BH with EBG gravity can be defined as
\begin{equation}
\Gamma=\frac{\Gamma_{emission}}{\Gamma_{absorption}}=
\exp\left[{-2\pi}\frac{(\tilde{E}-j\Omega-A_{0}e)}
{\kappa(r_{+})}\right]\left[1+\alpha\Xi\right],
\end{equation}
here
\begin{eqnarray}
\kappa(r_{+})&=&\frac{r_{+}^5\Big(\Upsilon-r_{+}^2\Big)-4Mr_{+}^4\beta+2a^2r_{+}\left(4q^2
\beta+r_{+}^4\Big(\Upsilon-r_{+}^2\Big)-10Mr_{+}\beta\right)}{4\pi \Upsilon\beta\Big(r_{+}^2+a^2\Big)},\\
\Upsilon&=&\sqrt{r_{+}^2-8q^2\beta+16Mr_{+}\beta}.\nonumber
\end{eqnarray}
In the presence of gravity terms, we compute the
$T'_{H}$ of the $4$-dimensional rotating charged BH in EGB gravity theory
by considering the factor
$\Gamma_{B}=\exp\left[(E-j\omega-eA_{0})/T'_{H}\right]$ (Boltzmann factor) as
\begin{eqnarray}
T'_{H}&=&\frac{r_{+}^5\Big(\Upsilon-r_{+}^2\Big)-4Mr_{+}^4\beta+2a^2r_{+}\left(4q^2
\beta+r_{+}^4\Big(\Upsilon-r_{+}^2\Big)-10Mr_{+}\beta\right)}{8\pi \Upsilon\beta\Big(r_{+}^2+a^2\Big)}\Big[1-\alpha\Xi\Big].\label{b1}
\end{eqnarray}
This solution shows that the $T'_{H}$
depends on the BH geometry. Also, it is alike form that of the other scalar and fermion particles.
The modified temperature depends on the correction parameter $\alpha$,
spin parameter $a$, BH charge $q$, arbitrary constant $\beta$, BH mass $M$ and BH radius $r_+$.
Quantum corrections decelerate the increase in $T'_{H}$ during the radiation phenomenon.
 These corrections make the radiation cease at some particular Hawking temperature, leaving the remnant mass. When this consideration holds, the temperature stops rising \cite{R23}
\begin{equation}
(M-dM)(1+\alpha\wp)\approx M,
\end{equation}
where $\omega=dM,~~\alpha_{0}(\frac{1}{M_{Planks}})=\alpha $ and $M_{Planks}=\omega$. Here, $\alpha_0$ and $M_{Planks}$ are represent the dimensionless parameter and Planck
mass and
quantum gravity effects for $\alpha_0 < 10^{5}$ in \cite{40, R24, R25}.

\section{Graphical $T'_{H}$ Analysis}

This section describes the stability conditions of rotating charged BH with
EBG gravity with the quantum gravity effects.
We analyze the physical behavior and stability of the corresponding BH by
plotting the graphs of corrected temperature $T'_{h}$
via horizon $r_+$ for fixed value of arbitrary parameter $\Xi=1$.
\begin{center}
\includegraphics[width=7.3cm]{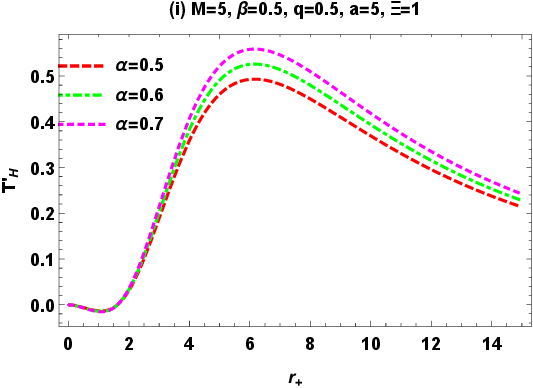}\includegraphics[width=7.3cm]{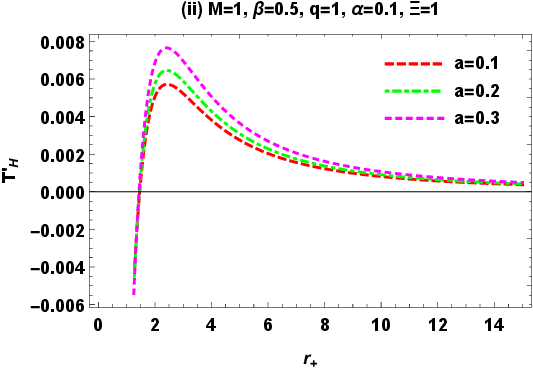}\\
{Figure 4: $T'_{H}$ via $r_{+}$ for $\Xi=1$ and $\beta=0.5$. Left $M=5=a$, $q=0.5$, $\alpha=0.5$ (dashed), $\alpha=0.6$ (dot-dashed), $\alpha=0.7$ (small-dashed). Right $M=1=q$, $\alpha=0.1$, $a=0.1$ (dashed), $a=0.2$ (dot-dashed), $a=0.3$ (small-dashed). }
\end{center}
Fig. \textbf{4} shows the corrected temperature $T'_H$ behavior w.r.t horizon $r_+$ under the effects of quantum gravity.

(i) describes the temperature behavior for different the correction
parameter $\alpha$ values and constant values of mass $M=5$, arbitrary parameter and charge $\beta=0.5=q$ and spin parameter $a=5$.
One cam observe the physical behavior (positive temperature) of BH under the effects
of correction parameter. Moreover, the $T'_H$ decreases as
$r_+$ increases, which also satisfies the Hawking's Phenomenon.

(ii) depicts the $T'_H$ behavior for constant values of mass $M=1$, arbitrary parameter
$\beta=0.5$, charge $q=1$, correction parameter $\alpha=0.1$ and different values of rotation parameter $a$.
At initial, the BH shows the non-physical behavior (negative temperature) but as time
runs out and after a maximum height the temperature slowly decreases as $r_+$ increases
as well as obtain an asymptotically flat condition to represents the complete stable form till $r_+ \rightarrow\infty$.

Fig. \textbf{5} represents the $T'_H$ behavior in the presence of correction
parameter $\alpha$ for fixed value of mass $M=1$ and in the absence of charge and spin
parameter ($q=0=a$), respectively.

(i) gives the temperature behavior in the absence of $q=0$ (charge) for constant values
of $\beta=0.5, a=10$ and different values of gravity parameter $\alpha$. One can observe that the temperature
gradually increases and then slowly decreases with the increasing $r_+$ that depicts the stable form of BH.

(ii) represents the $T'_H$ behavior in the absence of $a=0$ (spin parameter)
for fixed $\beta=0.1, q=5$ and for varying $\alpha$.
We can examine that the $T'_H$ behavior slowly decreases as horizon increases and gets an asymptotically flat condition that shows the stability of BH.

We can conclude from both plots that the BH is stable in the presence as well as absence
of charge and spin parameter under the influence of quantum gravity parameter effects.

\begin{center}
\includegraphics[width=7.9cm]{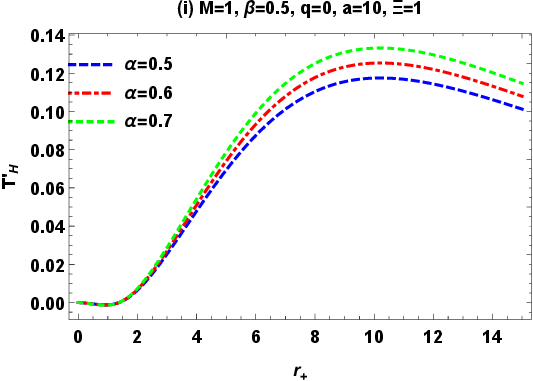}\includegraphics[width=7.9cm]{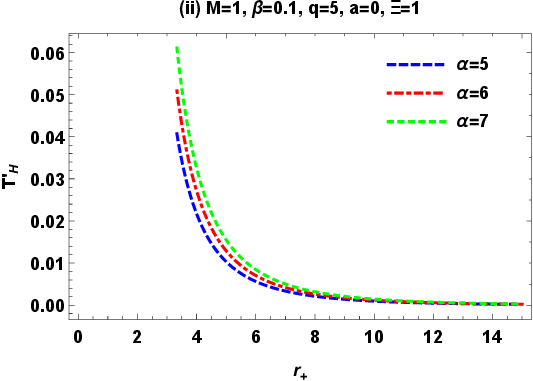}\\
{Figure 5: $T'_{H}$ via $r_{+}$ for $\Xi=1$ and $M=1$. Left $\beta=0.5$, $q=0$, $a=10$, $\alpha=0.5$ (dashed), $\alpha=0.6$ (dot-dashed), $\alpha=0.7$ (small-dashed). Right $\beta=0.1$, $q=5$, $a=0$, $\alpha=5$ (dashed), $\alpha=6$ (dot-dashed), $\alpha=7$ (small-dashed).}
\end{center}

\section{Logarithmic Correction of Entropy for Rotating Charged BH}

In this section, the Bekenstein-Hawking entropy is considered to be corrected by particular correction terms in the
quantum loop expansion. The logarithmic terms in BH entropy occur as a one-loop addition to the classical BH entropy. The BH entropy plays a significant role in the account of primordial BHs \cite{51}. It imposes a minimum mass of the primordial BHs so the small ones have enough time
to evaporate totally until the present form. The modified entropy with the logarithmic terms
demonstrate, this conclusion is no more in place and the BHs evaporation decompresses as soon as mass decreases below the critical
mass ($BH~~mass\simeq \sqrt{s_{0}}(planks~~mass)$) that extends to much longer BHs life time. In this form the small primordial BHs do not evaporate totally.
Now, we analyze the corrections
of entropy for EGB gravity of charged rotating BH.  Banerjee and Majhi \cite{47}-\cite{49} have studied
the corrections of Hawking temperature and entropy by taking into account the back-reaction effects via null
geodesic method. We calculate the corrections of entropy for EGB gravity of charged rotating BH by considering the Bekenstein-Hawking entropy formula for first order corrections \cite{50}.
We derive the logarithmic corrections of entropy by using
corrected temperature $T'_{H}$ and standard entropy $\mathbb{S}_{o}$ for EGB gravity of charged rotating BH.
The corrections of entropy can be computed by the following formula
\begin{equation}
\mathbb{S}=\mathbb{S}_{o}-\frac{1}{2}\ln\Big|T_H^2, \mathbb{S}_{o}\Big|+...~.\label{vv}
\end{equation}
The standard entropy for EGB gravity of rotating charged BH can be calculated by the following formula
\begin{equation}
\mathbb{S}_{o}=\frac{A_+}{4},
\end{equation}
here
\begin{eqnarray}
A_+&=&\int_{0}^{2\pi}\int_{0}^{\pi}\sqrt{g_{\theta\theta}g_{\phi\phi}}d\theta d\phi,\nonumber\\
&=&\frac{\pi\Big[2\Big(r_{+}^2+a^2\Big)^2-a^2q^2+r_{+}Ma^2\Big]}{\Big(r_{+}^2+a^2\Big)}.
\end{eqnarray}
So the entropy term for standard can be derived as
\begin{equation}
\mathbb{S}_{o}=\frac{\pi\Big[2\Big(r_{+}^2+a^2\Big)^2-a^2q^2+r_{+}Ma^2\Big]}{4\Big(r_{+}^2+a^2\Big)}.\label{v1}
\end{equation}
After substituting the values from Eq. (\ref{b1}) and (\ref{v1}) into Eq. (\ref{vv}),
we compute the correction of entropy as
\begin{eqnarray}
\mathbb{S}&=&\frac{\pi\Big[2\Big(r_{+}^2+a^2\Big)^2-a^2q^2+r_{+}Ma^2\Big]}{4\Big(r_{+}^2+a^2\Big)}\nonumber\\
&-&\frac{1}{2}\ln\left|\frac{\Big[\Big(r_{+}^5\Big(\Upsilon-r_{+}^2\Big)-4Mr_{+}^4\beta+2a^2r_{+}\left(4q^2
\beta+r_{+}^4\Big(\Upsilon-r_{+}^2\Big)-10Mr_{+}\beta\right)\Big)\Big(1-\alpha\Xi\Big)\Big]^2\chi}{256\pi \Upsilon^2\beta^2\Big(r_{+}^2+a^2\Big)^3}\right|+...,\label{b2}
\end{eqnarray}
where
\begin{equation*}
\chi=\Big[2\Big(r_{+}^2+a^2\Big)^2-a^2q^2+r_{+}Ma^2\Big].
\end{equation*}
The Eq. (\ref{b2}) represents the correction of entropy for EGB gravity of rotating charged BH.

\section{Conclusions}

In this article, we have used the Newman-Janis algorithm to study the charged rotating BH with EBG gravity
solution. We have derived the thermodynamics (Hawking temperature) at the outer
horizon for the corresponding BH by using a general formula for symmetric space-time.
The Hawking temperature depends upon rotation parameter $a$
(occurs due to Newman-Janis algorithm), BH mass $M$, BH charge $q$,
BH horizon $r_+$ and arbitrary constant $\beta$.
Moreover, we analyzed the graphical behavior of temperature w.r.t horizon
to analyze the stability of BH under the effects of spin parameter. We have also
investigated the stability of BH in the presence and absence of charge and spin parameter $q=0=a$, respectively.
We have also investigated the physical significance of our plots.
We conclude the main results from the graphical analysis of Hawking temperature via horizon as follows:
\begin{itemize}
\item For different values of spin parameter $a$ the $T_{H}$ slowly goes on increasing with the increasing values of horizon but after a height it gradually goes on decreasing with the increasing horizon which satisfies the Hawking’s phenomenon, so grantee’s the physical and stable form of BH. The temperature $T_{H}$ also increases with the increase in $a$.
\item For varying values of charge $q$ the BH at initial shows the non-physical form with negative temperature but as time goes on BH attains its stable form, when it eventually drops down from a height and obtains an asymptotically flat state till
$r_+ \rightarrow\infty$. The decreasing temperature with increasing horizon also shows the stable condition of BH. The temperature decreases with the increasing values of charge.
\item In the absence of spin parameter $a$, after a maximum height the $T_{H}$ behavior slowly decreases as horizon increases and gets an asymptotically flat condition  till
$r_+ \rightarrow\infty$ that shows the stability of BH.  From Fig. \textbf{2} and \textbf{3}, it is clear that in the presence of rotation parameter the temperature increases as compare in the absence of $a$.
\item In the absence of charge the temperature $T_{H}$ gradually increases and then slowly decreases with the increasing $r_{+}$ that depicts the stable form of BH. But it can be observed from the plots that BH is in more stable form in the absence of charge $q$.
\end{itemize}
In all plots, the BH depicts its stable condition by satisfying the Hawking's idea (the size of BH reduces when more radiations emit), the temperature is high at small horizon. We observe maximum temperature at minimum value of horizon.
Furthermore, we have calculated the Hawking temperature under
the influence of quantum gravity by using the semi-classical method and also examined the
graphical corrected temperature $T'_H$ behavior w.r.t horizon $r_+$ to check the physical
and stable conditions of BH under the effects of $\alpha$ (quantum gravity parameter). We also discussed the
stability conditions of BH in the absence of charge and spin parameter with or without gravity effects.
We have concluded that with or without quantum gravity the BH represents its stable form.
In our analysis we have found that the quantum corrections decelerate the increase in temperature during the
radiation process. This correction causes the radiation ceased at some specific
temperature, leaving the remnant mass. The remnant will obtain at specific condition
\begin{equation}
M_{Res} \simeq \frac{M_{p}^2}{\beta_0\omega}\gtrsim \frac{M_{p}}{\beta_0}.
\end{equation}

The results from the graphical analysis of corrected Hawking temperatures
with respect to the horizon for the given BH is summarized as
follows:
\begin{itemize}
\item The $T'_{H}$
decreases with the increasing horizon and BH reflects the stable state
for varying values of rotation parameter $a$ and correction parameter $\alpha$.
The corrected temperature $T'_{H}$ also increases with the increase in $a$ and $\alpha$.
The BH remnant can be obtained at nonzero horizon with maximum
temperature in the domain $0\leq r_{+}\leq15$.
\item In the absence of charge the corrected temperature $T'_{H}$ gradually increases and then slowly decreases with the increasing $r_{+}$ that depicts the stable form of BH.
\item In the absence of spin parameter $a$ the $T'_{H}$ behavior slowly decreases as horizon increases and gets an asymptotically flat condition  till $r_+ \rightarrow\infty$ that shows the stability of BH.
\end{itemize}
From our plots, we conclude that the tunneling emission rate increases (gives high temperature)
with the increasing values of $\alpha$. The corrected temperature of charged rotating BH with EBG gravity satisfies the both GUP and Hawking’s conditions, that guarantee the physical and stable states of BH.
According GUP condition the next order corrections must be
small as compared to the standard term of Uncertainty relation. The positive value of temperature in these plots
also satisfies the GUP relation, when GUP conditions does not satisfies temperature becomes negative (shows non-physical state of BH).
Now by comparing the plots in the presence/absence of quantum gravity parameter, we can observe that the
corrected temperature is lower that the original temperature.
In conclusion, the Newman-Janis algorithm gravity has attracted more attention for
different space-time. In this paper, we have evaluated the BH space-time
in Newman-Janis algorithm and computed the Hawking temperature with the effects of the rotation parameter
and correction parameter.
Finally, we computed the logarithmic corrected entropy for EGB gravity of rotating charged BH and also checked the logarithmic corrected entropy with the effects of the rotation parameter and correction parameter.

In future, we will study on the other
objects(i.e., wormholes and black rings) in the Newman-Janis algorithm.\\

\end{document}